\documentclass[aip, apl, 10pt, groupedaddress, onecolumn, floatfix, preprint]{revtex4-1}
\usepackage[T1]{fontenc}
\usepackage{bm}
\usepackage{graphicx}
\usepackage{amsmath}
\usepackage{amssymb}
\usepackage{color}
\usepackage{float}
\usepackage{ifpdf}

\newcommand{\eps}{\varepsilon}

\begin{document}
\title{\textbf{Strain field in graphene induced by nanopillar mesh}}

\author{S. P. Milovanovi\'{c}}\email{slavisa.milovanovic@uantwerpen.be}
\author{L. Covaci}\email{lucian@covaci.org}
\author{F. M. Peeters}\email{francois.peeters@uantwerpen.be}

\affiliation{Department of Physics, University of Antwerp \\
Groenenborgerlaan 171, B-2020 Antwerpen, Belgium}
\begin{abstract}
The mechanical and electronic properties of a graphene membrane placed on top of a superlattice of nanopillars are investigated. We use molecular dynamics (MD) simulations to access the deformation fields and the tight-binding approaches to calculate the electronic properties. The system of interest consists of a triangular lattice of nanopillars with a period of a=750 nm over which the graphene layer is deposited. Ripples form in the graphene layer that span across the unit cell, connecting neighbouring pillars, in agreement with  recent experiments. We find that the resulting pseudo-magnetic field (PMF) varies strongly across the unit cell. We investigate the dependence of PMF on unit cell boundary conditions, height of the pillars, and the strength of the van der Waals interaction between graphene and the substrate. We find direct correspondence with typical experiments on pillars, showing intrinsic "slack" in the graphene membrane. PMF values are confirmed by the local density of states calculations performed at different positions of the unit cell showing pseudo-Landau levels at varying spacings. Our findings regarding the relaxed membrane configuration and the induced strains are transferable to other flexible 2D membranes.
\end{abstract}


\date{Antwerp, \today}

\maketitle
\section{Introduction}
Recently, monolayer semiconductors gained a lot of interest due to their remarkable optical properties. It was shown that the monolayer transition-metal dichalcogenides (TMDs) have great potential as single-photon emitters. Numerous studies showed enhanced photoluminescence (PL) in these systems that originates from excitons trapped at the edges of TMD flakes \cite{r_pl01, r_pl02, r_pl03, r_pl04}. Namely, defects that appear in these systems (e.g. impurities, vacancies) serve as charge traps that locally change the potential of the sample and, thus, can be considered as quantum dots. It was shown that in the case of WSe$_2$ these dots are optically active and emit single photons \cite{r_pl02}. Single-photon emitters are of importance for the development of quantum information technology \cite{r_qip01, r_qip02, r_qip03}, 2D optoelectronics \cite{r_2do01,r_2do02}, novel metrology applications \cite{r_metro}, etc.

The above mentioned studies rely on naturally existing defects which means that the size and the position of the optically active regions cannot be controlled. This is not convenient from a practical point of view and a solution that is based on a more controllable technology is required. Strain fields proved to be a good alternative. Potentials generated in strained regions are strong enough to create quantum confinement and, similarly as in the case of defects, can be used as optically active regions. However, unlike the defect case, strain can be easily controlled which gives more freedom in designing reliable and reproducible devices. In a very popular configuration, a 2D material membrane is stretched over a mesh of nanopillars resulting in a network of optically active nanodots \cite{r_np00, r_np01, r_np02, r_np03, r_np04, r_np05, r_np06}. Furthermore, the absorption intensity varies with the radius, height, period, and the shape of the pillars \cite{r_np00, r_np03, r_np04, r_np06}.

Although, several studies have been carried out using 2D materials stretched over a mesh of nanopillars, little is known about the configuration and the strength of the pseudo-magnetic field induced by the strain. In this paper we examine in detail the influence of nanopillars on the mechanical and electronic properties of the atomically thin 2D materials placed over them. We use molecular dynamics (MD) simulations to access the deformation fields and implement tight-binding calculations to calculate the electronic properties. For practical reasons and because of already available experiments \cite{r_corr_reser, r_eva01}, we choose graphene as a material to be placed over the nanopillars, however, qualitative results obtained here can be applied to other 2D materials since the strain tensor and the pseudo-magnetic field are obtained from deformation fields which are not (or to a small extent when length scale of deformations is larger than the inter-atomic distance) dependent on the microscopic details of the material. The elastic properties of the material can be easily input as parameters in the coarse-grid mesoscopic model used for the membrane.

\section{MD simulations}
In order to simulate graphene sheets with experimentally relevant dimensions, we use a coarse mesoscale model. Atomistic simulations of the mechanical deformation are implemented through a classical molecular dynamics (MD) approach. We use the approach developed by Cranford and Buehler in Ref. \onlinecite{cranford2011}, which is able to model macroscopic elastic properties of graphene membranes. In this mesoscale model the total energy is expressed as
\begin{equation}
\label{md1}
E_{tot}=E_{stretch}+E_{shear}+E_{bend}+E_{adhesion},
\end{equation}
where $E_{stretch}$, $E_{shear}$, $E_{bend}$ and, $E_{adhesion}$ are the energies due to axial stretching, shear deformation, bending and van der Waals interaction with a substrate, respectively. For axial stretching, shear deformation and bending, simple harmonic potentials are used, with parameters that fit the elastic properties of graphene. The stretching energy is expressed as
\begin{equation}
\label{estretch}
E_{stretch}=\frac{1}{2}k_{t}(r-r_0)^2,
\end{equation}
where $k_t=470$ kcal mol$^{-1}$\AA$^{-2}$ is the spring constant and $r_0=2.5$ nm is the equilibrium distance between two neighboring grid points of a coarse triangular grid. The shear deformation energy is
\begin{equation}
\label{eshear}
E_{shear}=\frac{1}{2} k_{sh}(\phi-\phi_0)^2,
\end{equation}
where $k_{sh}=144$ kcal mol$^{-1}$rad$^{-2}$ is the spring constant and $\phi_0=\pi/3$ is the equilibrium in-plane angle of the triangular grid. Similarly, the bending energy can be written as
\begin{equation}
\label{ebend}
E_{bend}=\frac{1}{2} k_{bend}(\theta-\theta_0)^2,
\end{equation}
where $k_{bend}=16870$ kcal mol$^{-1}$rad$^{-2}$ is the spring constant and $\theta_0=\pi$ is the equilibrium out-of-plane angle. For the adhesion energy we use a van der Waals type potential with a distance parameter $\sigma_{LJ}=0.3$ nm and depth of the energy well $E^{vdW}=473$ kcal mol$^{-1}$ corresponding to the interaction between two graphene layers. This parameter is varied in our simulations in order to model the interaction with other types of substrates. The pillars are modelled as cylindrical objects which interact with the graphene membrane through a similar van der Waals type interaction.

To obtain the lowest energy configuration of the graphene membrane we perform highly efficient MD simulations on GPUs by using the HOOMD simulation software \cite{hoomd1,hoomd2} on a periodic system containing the mesoscale model of the graphene membrane interacting with a triangular pillar lattice and a planar substrate. The total energy is minimized with a relative tolerance criterion, $\delta E=10^{-11}$.  
\section{Tight-binding model}
\begin{figure*}[htbp]
\begin{center}\includegraphics[width=17cm]{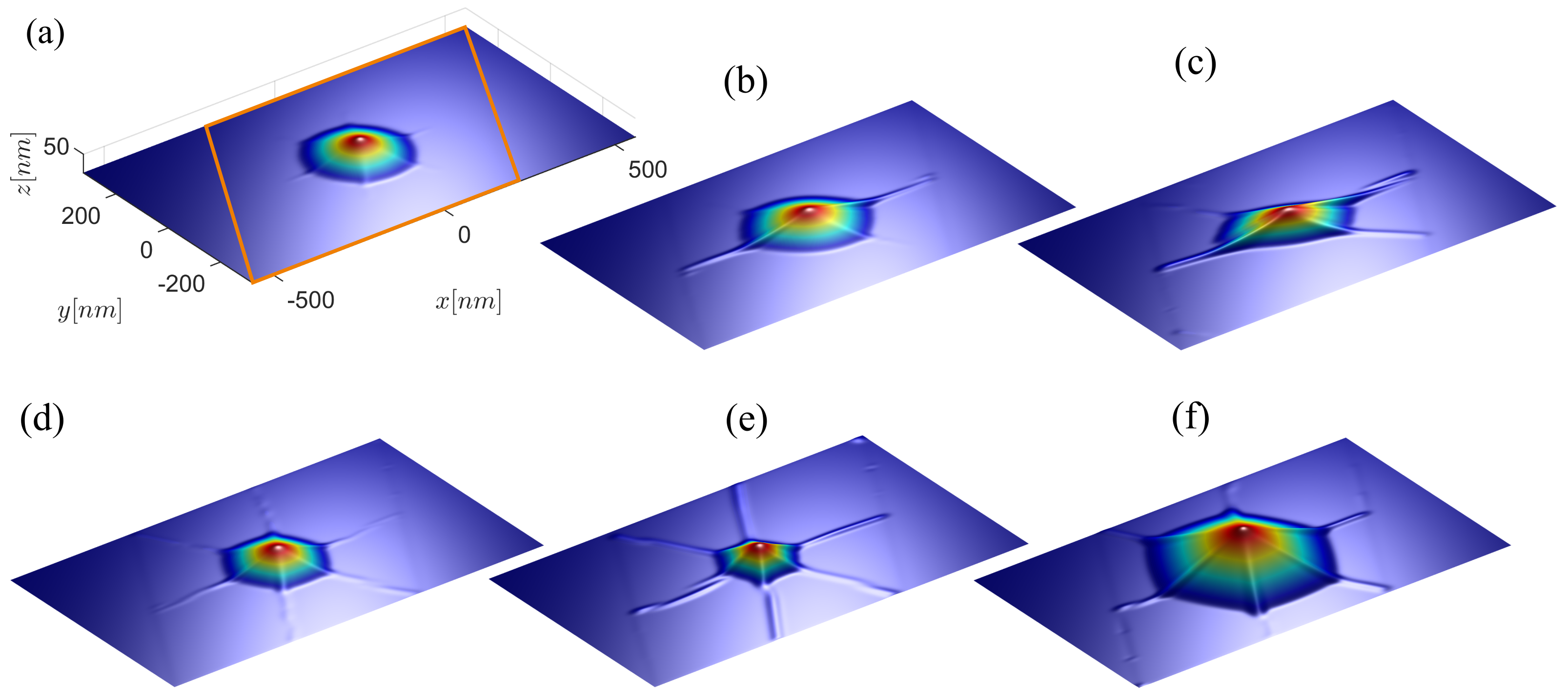}
\caption{MD simulations of graphene over a nanopillar with height $h=50$ nm. Plots are shown for different values of the compressibility of the unit cell (shown by orange rhombus) in the $x-$ and $y$-direction, $\varepsilon_x$ and $\varepsilon_y$. The following values are used: (a) $(\varepsilon_x$, $\varepsilon_y) = (0, 0)$, (b) $(0, 1\%)$, (c) $(0, 3\%)$, (d) $(1\%, 1\%)$, (e) $(3\%, 3\%)$, and (f) $(3\%, 3\%)$ with $h = 100$ nm.}
\label{fig_1}
\end{center}
\end{figure*}

To describe the electronic properties of strained graphene we use the standard single-band nearest-neighbor tight-binding Hamiltonian for the $\pi$ orbitals given by
\begin{equation}
\label{e_tb01}
H = \sum_{i}\epsilon_i c_{i}^{\dagger}c_i + \sum_{i,j} t_{ij}(\mathbf{r})c_{i}^{\dagger}c_j,
\end{equation}
where $c_{i}^{\dagger}$($c_i$) is the creation (annihilation) operator for an electron at site $i$, $\epsilon_i$ is the onsite potential at site $i$, and $t_{ij}$ is the hopping energy between sites $i$ and $j$. Notice that $t_{ij}$ is a spatially dependent function. Information about changes in the atomic positions extracted from MD simulations are plugged into Eq. \eqref{e_tb01} by changing the value of the hopping energy, which is modified according to the inter-atomic distance as
\begin{equation}
\label{e_tb02}
t_{ij}(\mathbf{r}) = t_0 e^{-\beta(r_{ij}/a_0 - 1)},
\end{equation}
where $t_0 = 2.8$ eV is the equilibrium hopping energy, $a_0=0.142$ nm is the length of the unstrained C-C bond, and $r_{ij} = \left| \mathbf{r}_i - \mathbf{r}_j\right|$ is the length of the strained bond between atoms $i$ and $j$. The decay factor $\beta = \partial \log t / \partial \log a \mid_{a=a_0} \approx 3.37$ describes how the hopping energy changes when the bond-length is modified \cite{r_tb01}. In the model, the zigzag edge of the graphene is aligned with the $x$-axis.

The Pybinding \cite{r_tb02} package was used to generate the tight-binding Hamiltonian and calculate local properties of the system. This easy-to-use python package is developed to simulate large systems that contain millions of atoms. Local properties of meso-systems are calculated using the kernel polynomial method (KPM) based on the Chebyshev expansion, an approximate method that is well suited for large systems and provides good agreement with methods based on exact diagonalization of the Hamiltonian \cite{r_kpm}. As such, it has been used in many problems in condensed matter physics \cite{r_kpm1, r_kpm2, r_kpm3, r_kpm4}.
\section{The deformation field}

\begin{figure}[b]
\begin{center}
\includegraphics[width=8.5cm]{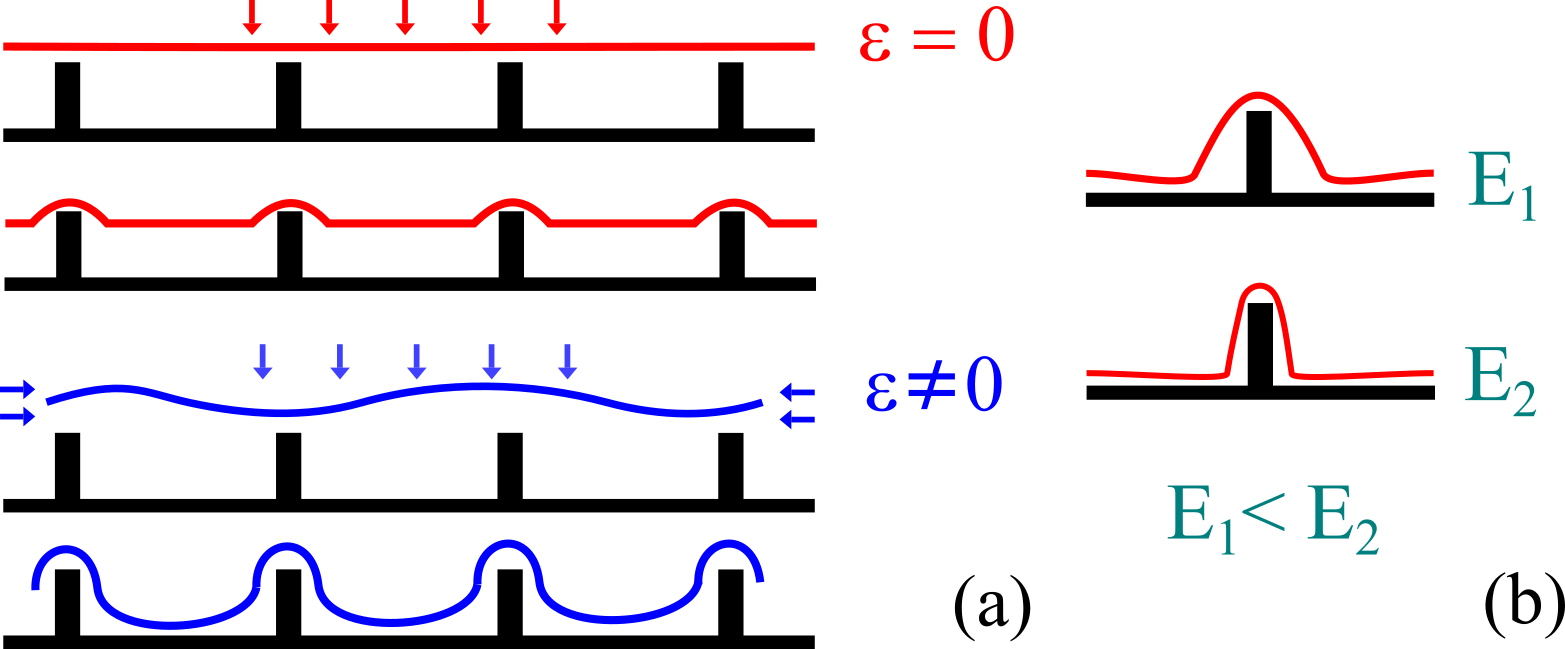}
\caption{(a) Cartoon drawing of graphene placed over the nanopillar mesh without and with pre-compression of the unit cell. (b) The increase of the van der Waals energy, $E^{vdW}$, leads to stronger interaction between the two materials, hence, the graphene layer wraps tighter around the pillar.}
\label{fig_2}
\end{center}
\end{figure}

We consider a graphene layer placed over the triangular lattice of nanopillars. The nanopillars are modeled as cylinders of radius $r$ (in the rest of the text we use $r = 10$ nm) and height $h$. The unit cell of such a system is shown in Fig. \ref{fig_1}(a) with an orange rhombus, using $h = 50$ nm. Notice that the dimensions of the unit cell, $\sim 750 \times 750$ nm, are chosen to reflect experimentally realizable systems. Here, we kept boundaries of the unit cell fixed, i.e. no compression of the unit cell is allowed (see top part of Fig. \ref{fig_2}(a)). The resulting out-of-plane deformation exhibits six-fold symmetry with noticeable ridges that stretch radially from the pillar towards the edges. Looking from the top, graphene forms a hexagonal pyramid around the pillar with each face separated by two ridges. This is quite different from the findings given in Ref. \onlinecite{r_np07} where the small size of the considered system ($\sim 35 \times 35$ nm) resulted with a smooth deformation profiles. However, experimental results showed that these findings are not valid for mesoscopic systems. In Ref. \onlinecite{r_eva01}, authors observed very pronounced ridges that stretch from one pillar to the neighboring one forming a strain network. Unfortunately, these results are also different from what we see in Fig. \ref{fig_1}(a) where the ridges are only observed around the pillars and disappear well before the boundaries of the unit cell. Thus, we need to find a way to enhance their presence. This is done by adding a small pre-compression to the unit cell of graphene, $\boldsymbol{\varepsilon}$, which induces wrinkling of graphene \cite{r_wri_01}.

 The idea behind it is shown in Fig. \ref{fig_2}(a). Placing the graphene layer over a mesh of nanopillars distributes the strain in graphene mostly around the pillar, while far away from the pillar, graphene is barely strained (top part of Fig. \ref{fig_2}(a)). Pillars serve as a support layer for graphene that hovers above the substrate layer. Distance between the graphene layer and the substrate layer depends on the parameters of the nanopillar mesh (e.g. density of pillars, their shape, height, structure, etc.) \cite{r_np00, r_np04, r_corr_reser, r_corr_2, r_corr_3}. However, when the unit cell of graphene is pre-compressed, as shown in the bottom part of Fig. \ref{fig_2}(a), graphene will have more "room" to envelope the pillar and  move towards the substrate. This is a more realistic scenario (compared to the case without pre-compression) since graphene has naturally rippled structure.  
 
 In Figs. \ref{fig_1}(b-f) we show how graphene's configuration changes when the values of pre-compression in the $x$- and $y$-direction, $\varepsilon_x$ and $\varepsilon_y$, change. First, one notices that the ridges are indeed enhanced by introducing small pre-compression. All figures show ridges that extend up to the edges of the unit cell. Furthermore, the effect of the pre-compression can be seen by comparing Figs. \ref{fig_1}(b-c) with Figs. \ref{fig_1}(d-f). In the first two figures, we pre-compress the unit cell only in the $y$-direction which means that graphene buckle along this axis. As a consequence, the ridges form along the $x$-axis. When the unit cell is pre-compressed in both directions equally, as shown in the bottom panel of Fig. \ref{fig_1}, the unit cell resembles the one from Fig. \ref{fig_1}(a). We again observe a hexagonal pyramid around the pillar with the difference that the ridges spread across the entire unit cell. By comparing these results with the experimental results from Ref. \onlinecite{r_eva01}, we can verify the existence of uniaxial strain in the sample from the geometry of the wrinkle network alone.
\begin{figure*}[t]
\begin{center}\includegraphics[width=17cm]{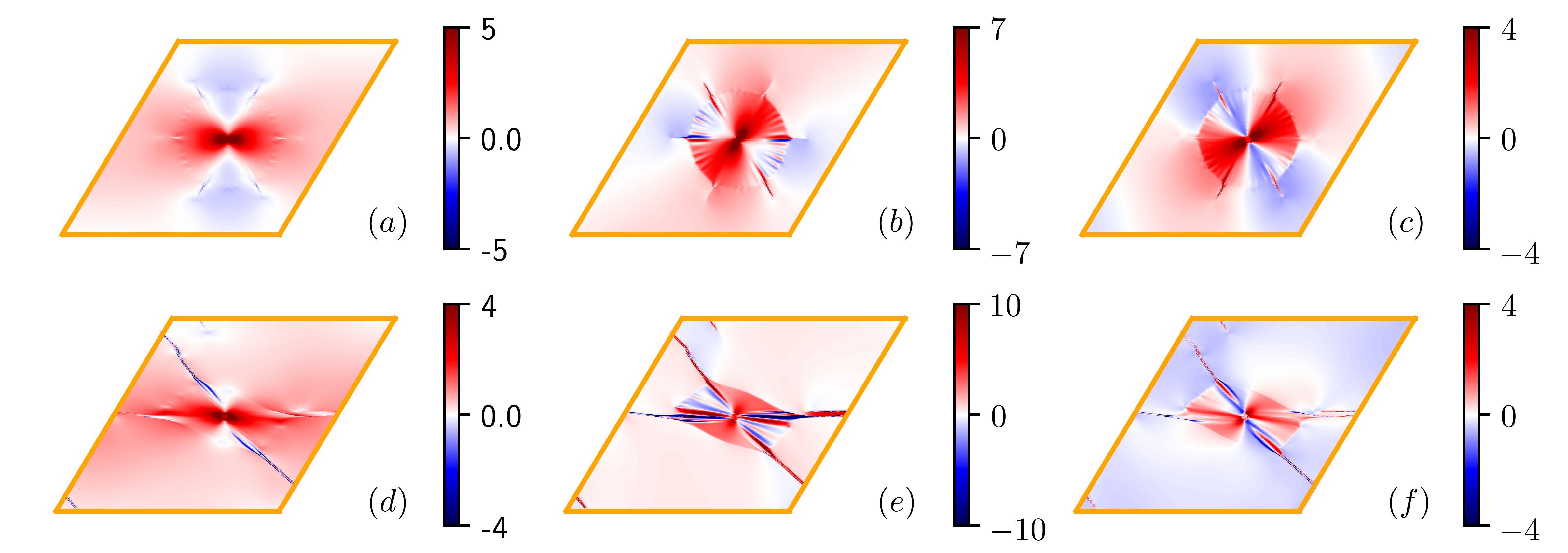}
\caption{Strain tensor elements: (a) $\varepsilon_{xx}$, (b) $\varepsilon_{yy}$, and (c) $\varepsilon_{xy}$ for the system from Fig. \ref{fig_1}(a). (d-f) The same as (a-c) but for the system from Fig. \ref{fig_1}(c).}
\label{fig_3}
\end{center}
\end{figure*}

The strain distribution is shown in Fig. \ref{fig_3}. Here, we plot elements of the strain tensor: $\varepsilon_{xx}$, $\varepsilon_{yy}$, and $\varepsilon_{xy}$ for the unit cells shown in Figs. \ref{fig_1}(a) and (c). The top part of Fig. \ref{fig_3} corresponds to the system shown in Fig. \ref{fig_1}(a), i.e. a configuration with no pre-compression of the unit cell. Strain components are comprised of stretched and compressed regions that form around the pillar. The largest deformation that graphene sustain is around the pillar itself where the strain goes up to 9$\%$. However, this is still well below the breaking limit for graphene\cite{r_stre_01}. Apart from this, large compression and stretching are observed around the ridges which is expected since this region exhibits the highest curvature. The bottom panel of Fig. \ref{fig_3} corresponds to the system shown in Fig. \ref{fig_1}(c), where the unit cell of graphene was pre-compressed in one direction and then the system is relaxed. Interestingly, strain around the pillar is lower in this case compared to the previous one and amounts to 5$\%$. This is not surprising, since in this case the system is pre-compressed. Regions with largest strain are now shifted towards the ridges where we observe large curvatures accompanied by a fast variation between stretching and compression of graphene. This is best illustrated in Figs. \ref{fig_3}(d) and (e) which show negative (positive) $\varepsilon_{xx}$ along the ridge, i.e. graphene is compressed (stretched), and positive (negative) $\varepsilon_{yy}$ at the same position. 

\begin{figure}[htbp]
\begin{center}
\includegraphics[width=6.5cm]{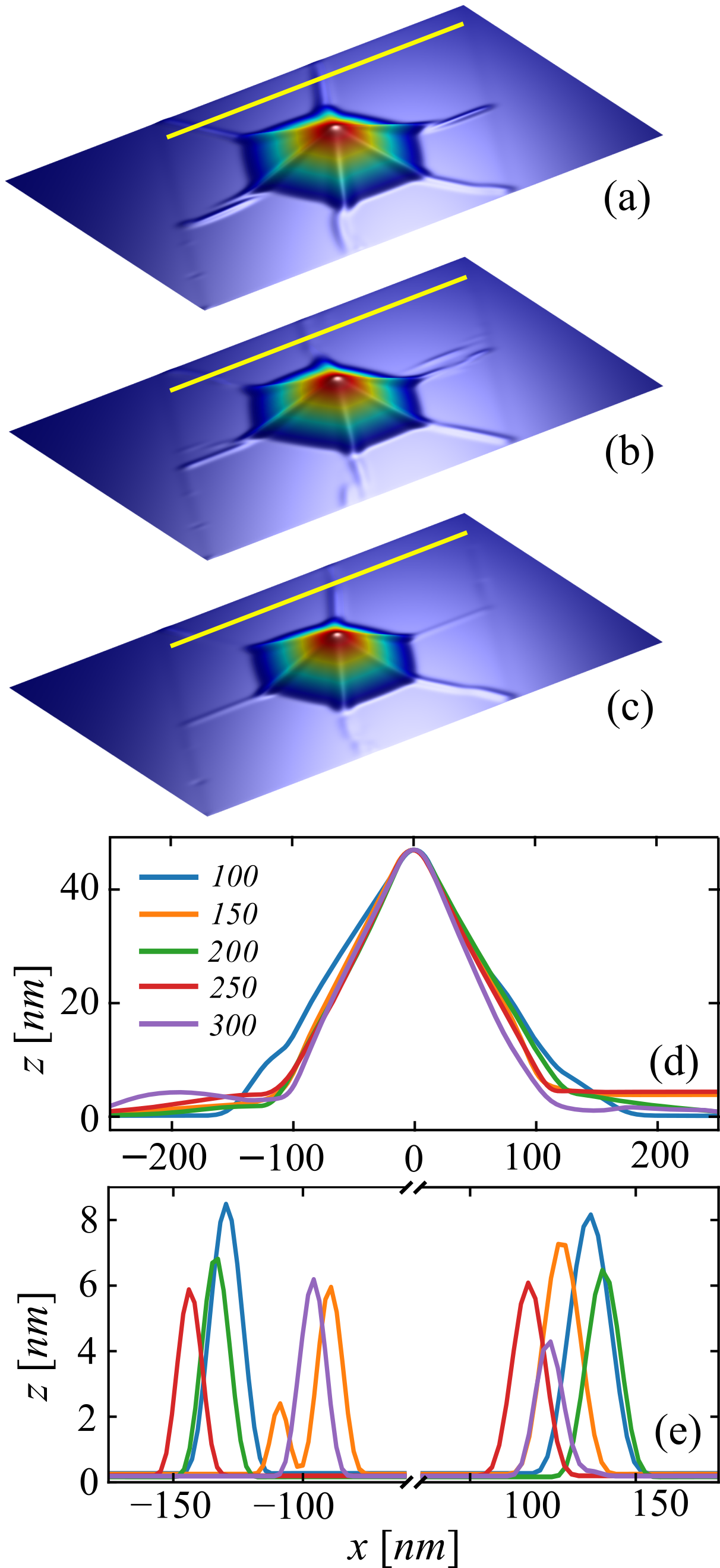}
\caption{Out-of-plane deformation for three different values of the van der Waals energy: (a) $E^{vdW} = 200$, (b) $E^{vdW} = 300$, and (c) $E^{vdW} = 400$ kcal mol$^{-1}$ using $h=50$ nm and $\pmb{\varepsilon} = (3 \%, 3\%)$. Cuts of $z$ along the (d) $y = 0$ and (e) $y = 220$ nm (yellow line in (a-c)) line for five different values of $E^{vdW}$ (in units of kcal mol$^{-1}$) shown in the inset.}
\label{fig_4}
\end{center}
\end{figure}

We next examine the effect of the van der Waals coupling between the membrane and the substrate on the deformation configuration. The results are shown in Fig. \ref{fig_4}. One can immediately notice from Figs. \ref{fig_4}(a-c) that an increase of $E^{vdW}$ leads to changes in the position and the shape of the ridges. A better insight in these changes can be obtained from Figs. \ref{fig_4}(d) and (e) where we plot cuts of $z$ at $y = 0$ and $y = 220$ nm, respectively. First plot shows the cut that crosses the pillar itself. Hence, it gives us information about the change of $z$ close to the pillar. We see that the increase in the interaction between the two layers is manifested by the narrowing of the out-of-plane displacement profile. Similar behaviour can be observed in Fig. \ref{fig_4}(e), where we plot the out-of-plane displacement along the $y = 220$ nm line (the yellow line in Figs. \ref{fig_4}(a-c)). We see that the general trend is that the ridges become narrower and smaller as $E^{vdW}$ increases. This is important because narrower features mean that the rate of change of graphene bond stretching is increased, suggesting an increase in the pseudo-magnetic field (PMF). This will be shown in the next section.
\section{The pseudo-magnetic field}
The spatial variation of the hopping energy is equivalent to the generation of a pseudo-magnetic vector potential, $\mathbf{A} = (A_x, A_y, 0)$, which can be evaluated around the $\mathbf{K}$ point using\cite{r_pmf1}
\begin{equation}
\label{e_vecpot}
A_x - \mathtt{i} A_y = -\frac{1}{ev_F}\sum_j \delta t_{ij} e^{\mathtt{i}\mathbf{K}\cdot \mathbf{r_{ij}}},
\end{equation}
where the sum runs over all neighboring atoms of atom $i$, $v_F$ is the Fermi velocity, and $\delta t_{ij} = (t_{ij} - t_0)$.
It is shown that using the linear expansion of Eq. \eqref{e_vecpot} one can easily connect the vector potential to the strain tensor $\pmb{\varepsilon}$ as \cite{r_linapp}
\begin{equation}
\label{ch8e23}
\mathbf{A} = -\frac{\hbar \beta}{2ea_{cc}} \begin{pmatrix}
\varepsilon_{xx} - \varepsilon_{yy} \\ -2\varepsilon_{xy}
\end{pmatrix},
\end{equation}
where $\varepsilon_{ij}$ are the elements of the tensor given by
\begin{equation}
\label{estx02}
\varepsilon_{ij} = \frac{1}{2} \left( \partial_j u_i +  \partial_i u_j + (\partial_iu_z)(\partial_ju_z) \right),~~~~~ i, j = x, y.
\end{equation}
The pseudo-magnetic field is then obtained as
\begin{equation}
\label{ch8e3}
\mathbf{B_{ps}} = \mathbf{\bigtriangledown} \times \mathbf{A} = \left(0,0,\partial_x A_y - \partial_y A_x \right)= \left(0,0,B_{ps}\right).
\end{equation}
It is important to mention that the PMF calculated for the  $\textbf{K'}$ point has the opposite direction compared to the one in the $\textbf{K}$ point.
\begin{figure*}[t]
\begin{center}\includegraphics[width=17cm]{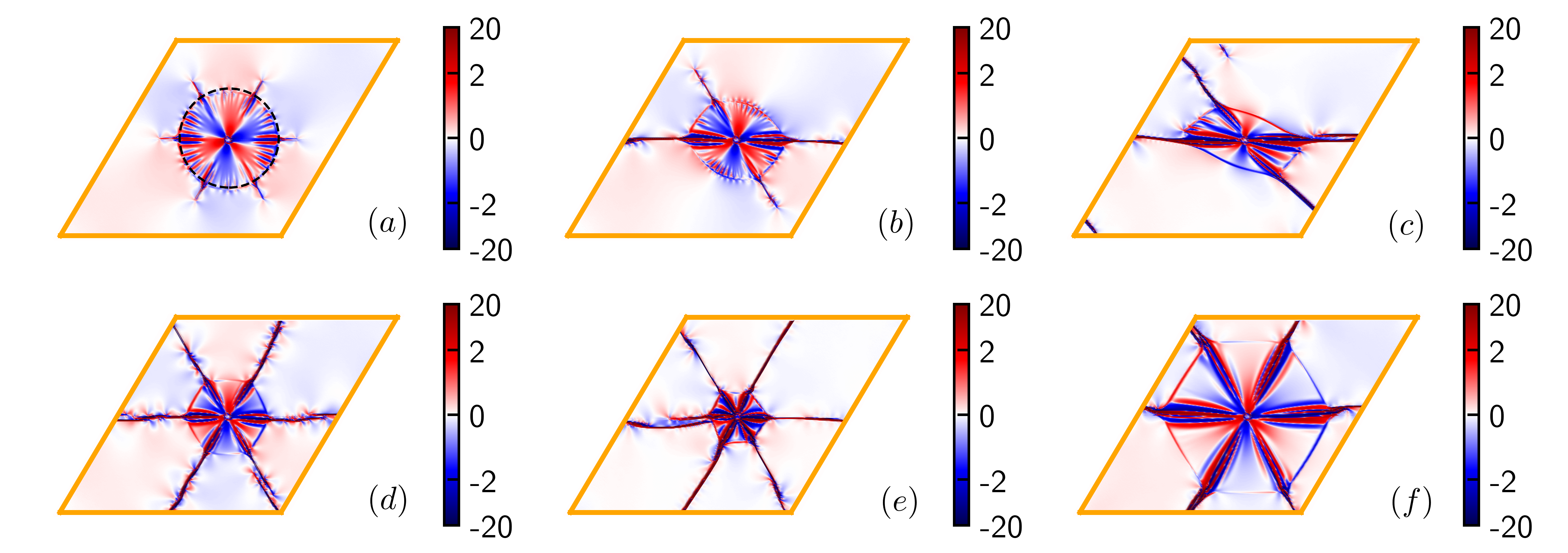}
\caption{(a-f) Profile of the pseudo-magnetic field for the displacements from Fig. \ref{fig_1}. Note that the color map is in logarithmic scale (units are in Tesla).}
\label{fig_5}
\end{center}
\end{figure*}

The pseudo-magnetic field profiles corresponding to the out-of-plane deformations shown in Fig. \ref{fig_1} are shown in Fig. \ref{fig_5}. When there is no pre-compression of the unit cell the PMF is 3-fold symmetric with alternating positive and negative regions, as shown in Fig. \ref{fig_5}(a). This profile resembles the ones obtained for Gaussian deformation \cite{r_gaus_01}, triangular strain \cite{r_tri_01}, and bubbles \cite{r_bubb_01}, which all show similar patterns. However, in this case regions of opposite direction of the PMF are separated by ridges, i.e. each face of the pyramid formed around the pillar shows PMF of the same sign. The exception is the radial region around the pillar where the out-of-plane displacement reaches its minimum. This region is marked by the dashed black circle in Fig. \ref{fig_5}(a). In the following we refer to this radius as $r_z$. Here, due to structural instabilities PMF shows fast variations in its sign and magnitude. Outside this region, generated PMF is significantly weaker due to the lower straining in this region, as shown in Figs. \ref{fig_3}(a-c). 

It is interesting to see the influence of the pre-compression of the unit cell on the pseudo-magnetic field profile. This is nicely illustrated in the top part of Fig. \ref{fig_5} where we show cases of $\varepsilon_{y}= 0$, $\varepsilon_{y}= 1 \%$, and $\varepsilon_{y}= 3 \%$, respectively. Notice how the PMF profile changes with pre-compression from a symmetric one with clearly distinguishable regions of different polarity in Fig. \ref{fig_5}(a) to a highly asymmetric PMF profile with fast changes of the PMF on each face of the out-of-plane displacement and strongly pronounced ridges that spread over the unit cell, shown in Fig. \ref{fig_5}(c). Note that along the ridges, the strain components vary fast, thus, due to the discreetness of the lattice calculation of the derivatives from Eq. \eqref{ch8e3} can result into unrealistically large values of the pseudo-magnetic field. The unit cell shown in Fig. \ref{fig_5}(b), represents an intermediate case. Here, the ridges are developed but the alternating regions of positive and negative magnetic field are still well defined. Although, unlike the case from Fig. \ref{fig_5}(a), here, not all the faces of the out-of-plane displacement have a PMF of the same sign. In the top (bottom) face of the tetragonal pyramid left (right) part has positive (negative) values of PMF, while right (left) part is negative.

Another interesting feature is shown in Figs. \ref{fig_5}(d-e). Here, we show two systems with different values of the initial equibiaxial compression of the unit cell. Both figures show similar features - six ridges that spread over the unit cell. However, observe how the hexagonal pyramid around the pillar shrinks with increase of the compression and the out-of-plane deformation becomes highly localized. These features agree well with the experimental data from Ref. \onlinecite{r_corr_reser} (see Fig. 2 of the main manuscript) for low density of the pillars. Similar observation can be made for the ridges - increase of compression leads to narrower ridges. As a consequence, we see that the PMF also becomes more localized. Fig. \ref{fig_5}(e) shows that the non-zero PMF is found only around the pillar and the ridges, unlike the case from Fig. \ref{fig_5}(d) where the non-zero field, although very weak, is found over the entire unit cell. Finally, in Fig. \ref{fig_5}(f) we plot a system with $h = 100$ nm, twice the size of the pillar from Fig. \ref{fig_5}(e). This figure gives us a nice view of the distribution of the PMF in the system with equibiaxial compression.
\begin{figure}[b]
\begin{center}
\includegraphics[width=8.5cm]{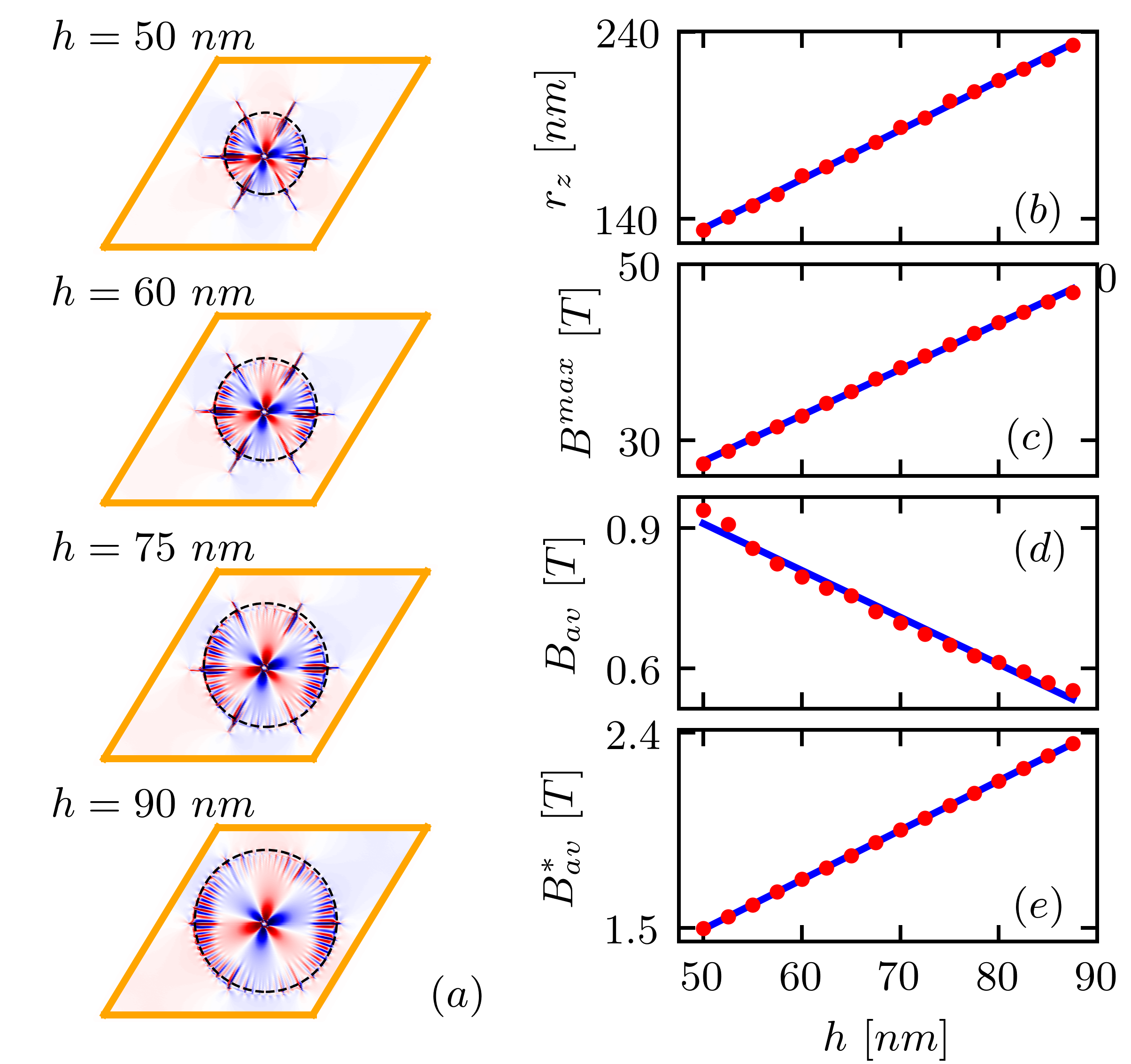}
\caption{(a) The PMF profile for different values of $h$ shown in the inset. Dashed circles indicate the radius at which the out-of-plane displacement reaches its minimum. The same color scale is used for all plots. (b) Change of the radius $r_z$ (explained in the main text) with $h$. (c) Change of the maximal value of the PMF with $h$. (d) The dependence of the average absolute value of the PMF over the radius $r_z$ around the pillar on the height of the pillar. (e) The same as (d) but now averaged over the radius of $r = r_z(h = 50$nm$)$ around the pillar.}
\label{fig_6}
\end{center}
\end{figure}

Up to now, we observed how the PMF changes with pre-compression of the unit cell. Now, we focus on the influence of pillar height on the pseudo-magnetic field. As seen from Figs. \ref{fig_5}(d-e) by increasing the height of the pillar we significantly increase the size of the "pyramid" around the pillar, i.e. the area with non-zero out-of-plane displacement. In order to study this in more detail, in Fig. \ref{fig_6}(a) we plot PMF for a few unit cells with different pillar heights using unit cell with no pre-compression. We choose this case because it is the easiest one to track the change of the PMF profile with $h$ due to its symmetry. In each plot, we mark the region of high PMF intensity with dashed circles, as we did in Fig. \ref{fig_5}(a). We confirm that the size of this region increases with $h$. In other words $r_z$ is a function of $h$. This dependence is plotted in Fig. \ref{fig_6}(b). The plot shows that $r_z$ is a linear function of the height of the pillar with a slope of approximately 2.6. Next, we investigate the change of the maximal value of the PMF with the height of the pillar, $B^{max}$. These results are shown in Fig. \ref{fig_6}(c). As $h$ is varied from $50$ nm to $90$ nm, $B^{max}$ linearly increases from $27$ T to $47$ T. However, due to the fast decay of the PMF in this region, $B^{max}$ is not a very useful quantity. Thus, in Fig. \ref{fig_6}(d) we plot the averaged absolute value of the PMF inside the area of radius $r_z$, $B_{av}$. One can see that this value decreases with increase of $h$. This is an unexpected result having in mind that $B^{max}$ almost doubles in the same range of pillar heights. However, if we look back to Fig. \ref{fig_6}(a), the reason for this behaviour becomes clearer. By increasing the height of the pillar, $r_z$ increases, however, the size of the region where the positive and negative PMF interchange does not follow this increase. Simply by comparing PMF for $h=50$ nm and $h = 90$ nm, one can observe that in the latter case PMF drops to zero much before reaching $r_z$ and consequently leading to a decrease of $B_{av}$. For this reason, we averaged the absolute value of the PMF in the region of constant radius of $r = r_z(h=50 $ nm$) = 134$ nm. These results are presented in Fig. \ref{fig_6}(e). As expected, new averaged value, $B_{av}^*$, indeed increases with $h$. Furthermore, by doubling the height of the pillars the pseudo-magnetic field in the region around the pillar almost doubles, as was the case with $B^{max}$. 
\begin{figure*}[t]
\begin{center}\includegraphics[width=17cm]{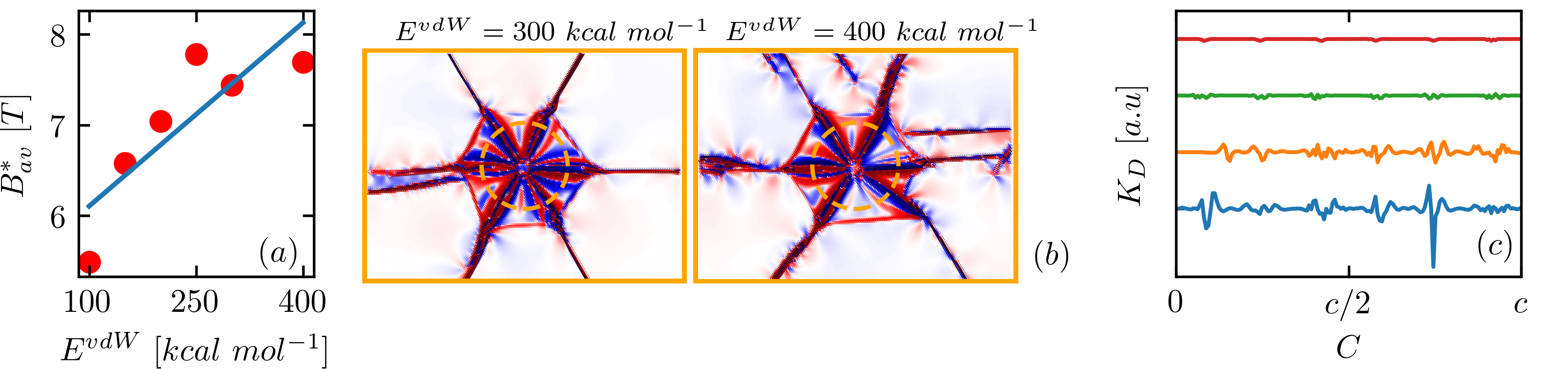}
\caption{(a) Average PMF in a region of radius $r = 120$ nm versus the van der Waals energy. (b) PMF profiles in regions around the pillar for two systems with $\eps_x = \eps_y = 3 \%$ and two values of the van der Waals energy given in the plot. The same color bar is used as in Fig. \ref{fig_5}. (c) Gaussian curvature, $K_D$, for the left (orange curve) and right (blue curve)  system from panels (b) calculated along the dashed line. The Gaussian curvature along the same radius for a system from Figs. \ref{fig_1}(a) (red curve) and (d) (green curve) is given for comparison. Curves are shifted for better visibility.}
\label{fig_6b}
\end{center}
\end{figure*}

As we mentioned in previous section, increasing the van der Waals energy leads to an increase of the curvature and consequently, of the pseudo-magnetic field. In Fig. \ref{fig_6b}(a) we plot the averaged absolute value of the PMF in a region of constant radius of 120 nm around the pillar versus the van der Waals energy, similarly as we did in Fig. \ref{fig_6}(e). Although, the general trend of increase of the PMF with $E^{vdW}$ is confirmed by this plot, we couldn't obtain a linear dependence, as we did in Fig. \ref{fig_6}(e). To understand this better, in Fig. \ref{fig_6b}(b) we plot two PMF profiles (in the zoomed region around the pillar) calculated for different values of the van der Waals energy. One can easily conclude that the ridges are not stable, or in other words, its position and continuity are very unpredictable. This strongly influences PMF, since the curvatures would change rapidly in this system. To confirm this we plot in Fig. \ref{fig_6b}(c) the Gaussian curvature along the dashed circle in Fig. \ref{fig_6b}(b) (orange and blue curve, respectively). Gaussian curvature is calculated using \cite{r_gc}
\begin{equation}
\label{egc}
K_D = (2\pi - \sum_i \theta_i)/A_p,
\end{equation}
where $\theta$ is the angle between two bonds and the sum goes over all nearest neighbours in the graphene lattice. $A_p$ is the area of the Voronoi tessellation. For comparison, we also added the Gaussian curvature along the same path for the two systems from Figs. \ref{fig_1}(a) and (d) (red and green curve, respectively). In the former case, curvature is rather large and changes quickly, as we expected. On the other hand, in the latter case, curvature is almost flat with the exception of the ridges, whose position can be easily extracted from this plot.
\begin{figure*}[t]
\begin{center}\includegraphics[width=14cm]{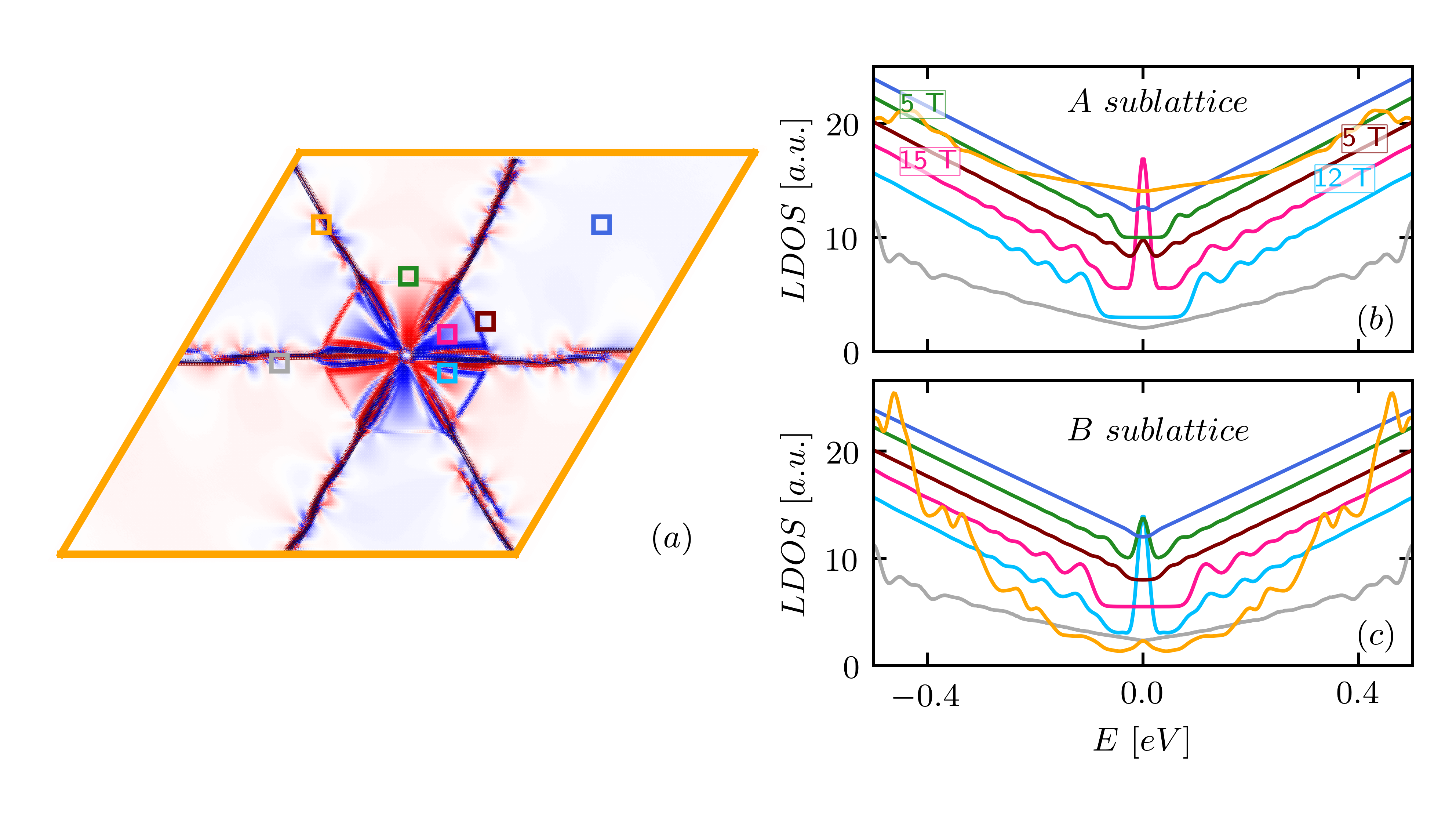}
\caption{(a) Pseudo-magnetic field for the system from Fig. \ref{fig_1}(d). Squares of different color show the points for which the LDOS is calculated. The LDOS dependence on energy is performed for both sublattices: (b) A and (c) B. Curves are shifted for better visibility.}
\label{fig_7}
\end{center}
\end{figure*}

Local density of states (LDOS) calculations for a system from Fig. \ref{fig_1}(d) are presented in Fig. \ref{fig_7}. Figs. \ref{fig_7}(b) and (c) show plots of LDOS versus the energy for two sublattices calculated for points marked by squares of different colors in Fig. \ref{fig_7}(a). Far away from the pillar, at a point marked by dark blue square, LDOS for both sublattices shows linear dependence on energy (dark blue curves in Figs. \ref{fig_7}(b) and (c)) confirming absence of (or very low) PMF in this region. Moving closer to the pillar but still outside the hexagonal pyramid (points marked by green and maroon square), the magnetic field increases enough for the pseudo-Landau levels to develop in the LDOS spectrum. The green and the maroon curve from Figs. \ref{fig_7}(b) and (c) tell us that the sign of the PMF in these two points is opposite since the zeroth pseudo-Landau level is observed at different sublattices. From the separation between the peaks in the LDOS we extract a pseudo-magnetic field of 5 T for these points. Coming even closer to the pillar, the pseudo-magnetic field increases. From the point marked with pink (light blue) square we extract 15 (12) T. Similarly as for the previous case, we conclude that these two points have pseudo-magnetic field in opposite directions. Finally, for points that belong to the ridges (gray and orange square) pseudo-Landau levels are not observed in the LDOS spectrum. However, plots show significant reduction of the LDOS in this region. As we already mentioned, due to the large curvatures in this region the pseudo-magnetic field is highly inhomogeneous around the ridges. Consequently, pseudo-Landau levels cannot develop in this region since a fairly constant field in a region of few magnetic lengths is needed for their appearance.
\section{Conclusions}
In this paper, we investigated the mechanical and electronic properties of graphene membranes deposited on a super-lattice of pillars. Our simulations showed that the deformation fields are very sensitive to the initial conditions applied to the unit cell. Adding small pre-compression or "slack" to the graphene membrane, changes the out-of-plane deformation completely and pronounced ridges appear in its structure forming a wrinkle network connecting the pillars. The size and the network configuration of the ridges can be used to determine the presence and the type of strain in the structure. 

Strain tensor elements showed that without pre-compression, the largest straining that graphene has to sustain is around the pillar itself where the strain reaches $\sim$10$\%$, still well below the breaking point for graphene. When the initial compression of the unit cell is considered, the region of largest strain moves to the ridges, followed by the fast variation of the curvature. Using different substrates does not change qualitatively our findings, i.e. the symmetry of the strain tensor elements and especially the ridges did not change. The change is rather quantitativly mirrored in a narrowing of the ridges with increase of the van der Waals energy and, generally, change the curvature due to the stronger interaction between the two materials.

Pseudo-magnetic field profiles revealed that around the pillar alternating regions of positive and negative PMF appear, mirroring the symmetry seen as obtained for Gaussian bumps or other circularly symmetric deformations. Especially, with no pre-compression, each face of the "pyramid" around the pillar has a PMF with only one direction. These regions start mixing when we apply an initial compression of the unit cell.

We studied the change of the parameters of the pseudo-magnetic field with the height of the pillars and the value of the van der Waals interaction. We found that in the case without pre-compression the maximal value of the PMF linearly increases with the height of the pillars. A linear dependence was also found for the average value of the PMF, taken over a radius of the area around the pillar with a non-zero out-of-plane displacement. The influence of the van der Waals energy is studied on the system with equibiaxial compression of $\eps_x = \eps_y = 3 \%$. Although, an increase of the $E^{vdW}$ results into an increase of the average value of the pseudo-magnetic field, the dependence is influenced by the fast change of the curvature in the region around the pillar.

Finally, the values of the PMF calculated from the deformation field are confirmed by our LDOS calculations performed using the tight-binding method. We found that on the faces of the pyramid, LDOS shows pseudo-Landau levels from which we extracted pseudo-magnetic fields of a few tens of Tesla. We expect that around the ridges the pseudo-magnetic field is much higher, however, this is not confirmed by the LDOS calculations. Possible reason is the fast oscillation of the field which prevents the Landau levels to develop.

Our findings can be, with slight modifications, applied to other 2D material membranes. Such systems are of interest in quantum technology as they were shown to be efficient single-photon emitters resulting from strain induced changes in the exciton confinement. Hence, our study can be used as a guide to optimize parameters of the system for the improvement of the confinement potential and, consequently, the efficiency of the emitters.
\section{Acknowledgment}
One of us (SPM) is supported by the Flemish Science Foundation (FWO).

\begin{thebibliography}{99}
%
%
%
\bibitem{r_pl01} P. Tonndorf, R. Schmidt, R. Schneider, J. Kern, M. Buscema, G. A. Steele, A. Castellanos-Gomez, H. S. J. van der Zant, S. M. de Vasconcellos, and R. Bratschitsch, Optica \textbf{2}, 347 (2015).
%
\bibitem{r_pl02} C. Chakraborty, L. Kinnischtzke, K. M. Goodfellow, R. Beams, and A. N. Vamivakas, Nat. Nanotech. \textbf{10}, 507 (2015).
%
\bibitem{r_pl03} M. Koperski, K. Nogajewski, A. Arora, V. Cherkez, P. Mallet, J.-Y. Veuillen, J. Marcus, P. Kossacki, and M. Potemski, Nat. Nanotech. \textbf{10}, 503 (2015).
%
\bibitem{r_pl04} A. Srivastava, M. Sidler, A. V. Allain, D. S. Lembke, A. Kis, and A. Imamo\u{g}lu, Nat. Nanotech. \textbf{10}, 491 (2015).
%
\bibitem{r_qip01} A. Imamo\u{g}lu, D. D. Awschalom, G. Burkard, D. P. DiVincenzo, D. Loss, M. Sherwin, and A. Small, Phys. Rev. Lett. \textbf{83}, 4204 (1999).
%
\bibitem{r_qip02} D. Press, T. D. Ladd, B. Zhang, and Y. Yamamoto, Nature \textbf{456}, 218 (2008).
%
\bibitem{r_qip03} S. G. Carter, T. M. Sweeney, M. Kim, C. Soo Kim, D. Solenov, S. E. Economou, T. L. Reinecke, L. Yang, A. S. Bracker, and D. Gammon, Nat. Photon. \textbf{7}, 329 (2013).
%
\bibitem{r_2do01} F. H. L. Koppens, T. Mueller, Ph. Avouris, A. C. Ferrari, M. S. Vitiello,  and M. Polini, Nat. Nanotech. \textbf{9}, 780 (2014).
%
\bibitem{r_2do02} S. Chander Dhanabalan, J. S. Ponraj, H. Zhang, and Q. Bao, Nanoscale \textbf{8}, 6410 (2016).
%
\bibitem{r_metro} A. N. Vamivakas, Y. Zhao, S. F\"{a}lt, A. Badolato, J. M. Taylor, and M. Atat\"{u}re, Phys. Rev. Lett. \textbf{107}, 166802 (2011).
%
\bibitem{r_np00} Z. Fan, R. Kapadia, P. W. Leu, X. Zhang, Y.-L. Chueh, K. Takei, K. Yu, A. Jamshidi, A. A. Rathore, D. J. Ruebusch, M. Wu, and A. Javey, Nano Lett. \textbf{10}, 3823 (2010).
%
\bibitem{r_np01} W. Li, S. Wang, S. He, J. Wang, Y. Guo, and Y. Guo, Sci. Rep. \textbf{5}, 11375 (2015).
%
\bibitem{r_np02} N. V. Proscia, Z. Shotan, H. Jayakumar, P. Reddy, M. Dollar, A. Alkauskas, M. Doherty, C. A. Meriles, V. M. Menon, Optica \textbf{5}, 1128 (2018).
%
\bibitem{r_np03} C. Palacios-Berraquero, D. M. Kara, A. R.-P. Montblanch, M. Barbone, P. Latawiec, D. Yoon, A. K. Ott, M. Loncar, A. C. Ferrari, and M. Atat\"{u}re, Nat. Commun. \textbf{8}, 15093 (2017).
%
\bibitem{r_np04} Y. Liu, W. Huang, T. Gong, Y. Su, H. Zhang, Y. He, Z. Liu, and B. Yu, Nanoscale \textbf{9}, 17459 (2017).
%
\bibitem{r_np05} M. Nguyen, S. Kim, T. Trong Tran, Z.-Q. Xu, M. Kianinia, M. Toth, and I. Aharonovich, Nanoscale \textbf{10}, 2267 (2018).
%
\bibitem{r_np06} A. Branny, S. Kumar, R. Proux, and B. D. Gerardot, Nat. Commun. \textbf{8}, 15053 (2017).
%
\bibitem{r_corr_reser} A. Reserbat-Plantey, D. Kalita, Z. Han, L. Ferlazzo, S. Autier-Laurent, K. Komatsu, C. Li, R. Weil, A. Ralko, L. Marty, S. Gu\'{e}ron, N. Bendiab, H. Bouchiat, and V. Bouchiat, Nano Lett. \textbf{14}, 5044 (2014).
%
\bibitem{r_eva01} Y. Jiang, J. Mao, J. Duan, X. Lai, K. Watanabe, T. Taniguchi, and E. Y. Andrei, Nano. Lett. \textbf{17}, 2839 (2017).
%
\bibitem{cranford2011} S. Cranford and M. J. Buehler, Model. and Sim. in Mat. Sci. and Eng. \textbf{19}, 054003 (2011).
%
\bibitem{hoomd1} J. Glaser, T. D. Nguyen, J. A. Anderson, P. Lui, F. Spiga, J. A. Millan, D. C. Morse, and S. C. Glotzer, Comp. Phys. Comm. {\bf 192}, 97 (2015).
%
\bibitem{hoomd2} J. A. Anderson, C. D. Lorenz, and A. Travesset, J. Comp. Phys. {\bf 227}, 5342 (2008).
%
\bibitem{r_tb01} V. M. Pereira, A. H. Castro Neto, and N. M. R. Peres, Phys. Rev. B \textbf{80}, 045401 (2009).
%
\bibitem{r_tb02} D. Moldovan, M. Andelkovic, and F. M. Peeters, Pybinding, DOI: 10.5281/zenodo.826942 (2017).
%
\bibitem{r_kpm} D. Moldovan, Ph.D. thesis, University of Antwerp, 2016 (\url{www.uantwerpen.be/en/research-groups/cmt/research/theses/}).
%
\bibitem{r_kpm1} H. R\"{o}der, R. N. Silver, D. a. Drabold, and J. J. Dong, Phys. Rev. B \textbf{55} 15382 (1997).
%
\bibitem{r_kpm2} L. Covaci, F. M. Peeters, and M. Berciu, Phys. Rev. Lett. \textbf{105}, 167006 (2010).
%
\bibitem{r_kpm3} J. H. Garcia, L. Covaci, and T. G. Rappoport, Phys. Rev. Lett., \textbf{114}, 116602 (2015).
%
\bibitem{r_kpm4} M. An\dj elkovi\'{c}, L. Covaci, and F. M. Peeters, Phys. Rev. Materials \textbf{2}, 034004 (2018).
%
\bibitem{r_np07} M. Neek-Amal, L. Covaci, and F. M. Peeters, Phys. Rev. B \textbf{86}, 041405(R) (2012).
%
\bibitem{r_wri_01} S. Deng and V. Berry, Mater. Today \textbf{19}, 197 (2016).
%
\bibitem{r_corr_2} J. H. Hinnefeld, S. T. Gill, and N. Mason, Appl. Phys. Lett. \textbf{112}, 173504 (2018).
%
\bibitem{r_corr_3} G. Li, C. Yilmaz, X. An, S. Somu, S. Kar, Y. Joon Jung, A. Busnaina, and K.-T. Wan, J. Appl. Phys. \textbf{113}, 244303 (2013).
%
\bibitem{r_pmf1}  V. M. Pereira and A. H. Castro Neto, Phys. Rev. Lett. \textbf{103}, 046801 (2009).
%
\bibitem{r_linapp} M. Vozmediano, M. Katsnelson, and F. Guinea, Phys. Reports \textbf{496}, 109 (2010).
%
\bibitem{r_stre_01} C. Lee, X. Wei, J. Kysar, and J. Hone, Science \textbf{321}, 385 (2008).
%
\bibitem{r_gaus_01} D. Moldovan, M. Ramezani Masir, and F. M. Peeters, Phys. Rev. B \textbf{88}, 035446 (2013).
%
\bibitem{r_tri_01} F. Guinea, M. I. Katsnelson, and A. K. Geim, Nat. Phys. \textbf{6}, 30 (2010).
%
\bibitem{r_bubb_01} K. Yue, W. Gao, R. Huang, and K. M. Liechti, J. Appl. Phys. \textbf{112}, 083512 (2012).
%
\bibitem{r_gc} A. A. Pacheco Sanjuan, Z. Wang, H. Pour Imani, M. Vanevi\'{c}, and S. Barraza-Lopez, Phys. Rev. B \textbf{89}, 121403(R) (2014).
%
%
%
\end{thebibliography}
\end{document}